\documentclass[prl,onecolumn,floatfix,superscriptaddress,a4paper,nofootinbib]{revtex4}
\usepackage{amsmath,bm,graphicx}
\usepackage{amssymb}
\usepackage{amsfonts}
\usepackage{epstopdf}
\usepackage{mathrsfs}
\usepackage{color}
\usepackage{latexsym}
\usepackage{hyperref}
\usepackage{subfigure}
\usepackage{bm}
\usepackage{natbib}
\usepackage{babel}

\usepackage[version=4]{mhchem}
\usepackage{booktabs}

\usepackage[nogroupskip,nonumberlist]{glossaries}
\setacronymstyle{long-short}
\newacronym{se}{SE}{Stokes-Einstein}
\newacronym{ses}{SES}{Stokes-Einstein-Sutherland}
\newacronym{svs}{SvS}{Svedberg-Stokes}
\newacronym{gk}{GK}{Green-Kubo}
\newacronym{pmf}{PMF}{potential of mean force}
\newacronym{com}{COM}{center of mass}
\newacronym{msd}{MSD}{mean square displacement}
\newacronym{rdf}{RDF}{radial distribution function}
\newacronym{acf}{ACF}{autocorrelation function}
\newacronym{vacf}{VACF}{velocity autocorrelation function}
\newacronym{facf}{FACF}{force autocorrelation function}
\newacronym{vdw}{vdW}{van der Waals}
\newacronym{fdt}{FDT}{fluctuation-dissipation theorem}
\newacronym{md}{MD}{molecular dynamics}
\newacronym{lj}{LJ}{Lennard-Jones}
\newacronym{wca}{WCA}{Weeks-Chandler-Andersen}
\newacronym{pme}{PME}{Particle mesh ewald}
\newacronym{NH}{NH}{Nos\'{e}-Hoover}
\newacronym{PR}{PR}{Parrinello-Rahman}
\newacronym{opls}{OPLS-AA}{Optimized Potentials for Liquid Simulations (all atoms)}
\newacronym{auc}{AUC}{analytical ultracentrifugation}
\newacronym{sfrs}{SFRS}{Soret forced Rayleigh scattering}
\newacronym{nmr}{NMR}{nuclear magnetic resonance}
\newacronym{lc}{LC}{liquid chromatography}
\newacronym{hplc}{HPLC}{high-performance liquid chromatography}
\newacronym{rplc}{RPLC}{reversed phase liquid chromatography}
\newacronym{hrtem}{HRTEM}{high resolution transmission electron microscopy}
\newacronym{stm}{STM}{scanning tunnelling microscopy}
\newacronym{xrd}{XRD}{X-ray diffraction}
\newacronym{pls}{PLS}{polystyrene latex particles}

\usepackage{siunitx}  

\renewcommand{\d}{\mathrm{d}}

\newcommand{\kBT}{k_\textup{B}T}

\newcommand{\kbt}{\kBT}
\newcommand{\kbT}{\kBT}

\newcommand{\ab}[1]{\textcolor{black}{#1}}

\begin{document}

\title{Modelling diffusive transport of particles interacting with \ab{slit} nanopore walls: The case of fullerenes in toluene filled alumina pores}

\author{Andreas Baer}
\affiliation{Friedrich-Alexander-Universität Erlangen-Nürnberg, Department of Physics, PULS Group, Interdisciplinary Center for Nanostructured Films (IZNF),Cauerstr. 3,91058 Erlangen, Germany}
\email[]{andreas.baer@fau.de}
\author{Paolo Malgaretti}
\affiliation{Helmholtz Institute Erlangen-N\"urnberg for Renewable Energy (IEK-11), Forschungszentrum J\"ulich, Cauerstr. 1,91058 Erlangen, Germany}
\email[Corresponding author: ]{p.malgaretti@fz-juelich.de }
\author{Malte Kaspereit}
\affiliation{Friedrich-Alexander-Universität Erlangen-Nürnberg, Department of Chemical and Biological Engineering, Institute for Separation Science \& Technology, Egerlandstrasse  3,      91058 Erlangen, Germany}
\email[]{malte.kaspereit@fau.de}
\author{Jens Harting}
\affiliation{Helmholtz Institute Erlangen-N\"urnberg for Renewable Energy (IEK-11), Forschungszentrum J\"ulich, Cauerstr. 1,91058 Erlangen, Germany}
\affiliation{Friedrich-Alexander-Universität Erlangen-Nürnberg, Department of Chemical and Biological Engineering and Department of Physics, Cauerstr. 1, 91058 Erlangen, Germany}
\email[]{j.harting@fz-juelich.de}
\author{Ana-Sun\v{c}ana Smith}
\affiliation{Friedrich-Alexander-Universität Erlangen-Nürnberg, Department of Physics, PULS Group, Interdisciplinary Center for Nanostructured Films (IZNF),Cauerstr. 3,91058 Erlangen, Germany}
\affiliation{Group of Computational Life Sciences, Department of Physical Chemistry, Ru\dj er Bo\v skovi\' c Institute, Bijeni\v{c}ka 54,10000 Zagreb,Croatia}
\email[Corresponding author: ]{smith@physik.fau.de, asmith@irb.hr}

\begin{abstract}
Accurate modeling of diffusive transport of nanoparticles across nanopores is a particularly challenging problem. The reason is that for such narrow pores the large surface-to-volume ratio amplifies the relevance of the nanoscopic details and of the effective interactions at the interface with pore walls.
Close to the pore wall, there is no clear separation between the length scales associated with molecular interactions, layering of the solvent at the interface with the pore and the particle size.
Therefore, the standard hydrodynamic arguments may not apply and alternative solutions to determining average transport coefficients need to be developed.
We here address this problem by offering a multiscale ansatz that uses effective potentials determined from molecular dynamics simulations to parametrise a four state stochastic model for the positional configuration of the particle in the pore.
This is in turn combined with diffusivities in the centre of the pore and at the pore wall to calculate the average diffusion constant.
We apply this model to the diffusion of fullerenes in a toluene filled \ab{slit} nanopore and calculate the mean diffusion coefficient as a function of the pore size.
We show that the accuracy of our model is affected by the partial slip of the toluene on the pore wall.   
\end{abstract}
\maketitle

\section{Introduction}
The transport of suspended particles through relatively narrow pores is important in a wide range of 
situations, examples of which include oil recovery~\cite{Foroozesh2020}, sap transport in plants~\cite{Jensen2016}, lymphatic flow~\cite{Moore2018}\ab{,} nanosensing devices \cite{Varongchayakul2018, Xue2020, Marion2021}
\ab{and devices relying on electroosmotic flow~\cite{Jubery2012, Li2019, Antaw2021}}.
From the point of view of daily applications, transport in pores is at the foundation of particle separation techniques, such as liquid chromatography \cite{Nicoud2015,Reithinger2011,Michaud2021}. It is therefore natural that a significant scientific effort has been invested to understand diffusive transport, also from the theoretical perspective~\cite{Hummer2005,Malgaretti2013,Elwinger2017,Coasne2016,Boccardo2020}.

Particle diffusion through \ab{(slit)} nanopores involves several characteristic time and length scales, spanning from the nanoscopic molecular interactions up to the mesoscopic pore size, and typically is described empirically by means of macroscopic models \cite{Thommes2010,DenUijl2021, Chen2022}.
These models account for the particle-pore interactions by means of effective potentials extracted from adsorption \linebreak[5]
isotherms~\cite{Thommes2010}  or Hamaker constants~\cite{Liu_2022}, as well as effective diffusion coefficients \cite{Bellot1991,Bellot1993}.
This is successful when the particle size is clearly much larger than the scale of viscosity gradients in the fluid close to the pore wall, or the range of the interaction potentials between the pore and the particle.
However, for nanoparticles suspended in organic solvents, there is no length scale separation.
Accordingly, the details of the atomistic interactions between the nanoparticles, the solvent, and the solid pore material are of enormous importance and cannot be accounted for by means of traditionally taken effective parameters. 
On the nanoscale, one established approach to studying porous structures are molecular dynamics (MD) simulations.
The latter are particularly useful for the quantification of basic interactions and dynamic processes determining adsorption or diffusion coefficients \cite{Klatte1996, Slusher1999, Lindsey2013, Rybka2016}.
So far, many pore types including biological, slit and round pores with different solvents  have been addressed \cite{Duo2006, Zhao2008, Lindsey2013, Melnikov2013, Rybka2019, Puza2022}.
However, when diffusive transport is strongly affected by adsorption and desorption from the pore wall, molecular simulations reach limits in the types of systems and solute concentrations that can be covered explicitly \cite{Lindsey2013}.

The limitations in sampling efficiency of unconstrained MD simulations can, for the interaction potential, be overcome by Monte Carlo simulations.
However, the latter do not include real time dynamics like diffusion.
Current approaches on incorporating static interactions and diffusive dynamics into a combined scheme for predicting larger-scale dynamics reduce the molecular results into effective diffusion coefficients and still do not account for particle-particle interactions \cite{Tallarek2019, Tallarek2022}.
It is therefore necessary to develop multiscale approaches that incorporate the molecular details in the study of diffusion, in order to provide accurate information on transport on the mesoscale. 

In this study, we propose a simple model to calculate the transport coefficients for nanoparticles in pores, by evaluating the interaction potentials between the particle and the pore, as well as particle-particle interactions in bulk and at the interface, where we also evaluate the diffusivity of particles. We then separate the space into an interfacial and a bulk region and calculate the probabilities to find the particle as a freely diffusing entity or as part of the complex. Convoluting this information with the appropriate diffusivities yields the effective diffusion coefficient, as a function of the pore width. We expect that this approach is particularly useful in the regime of strong particle-wall interactions and pore widths in which the opposing walls are sufficiently far such that there is bulk solvent in the central part of the pore. Consequently, our model accounts for solvation properties and the particle-particle interactions and predicts the retention times directly from a limited and accessible simulation effort.  

\begin{figure}
    \centering
    \includegraphics[width=\textwidth]{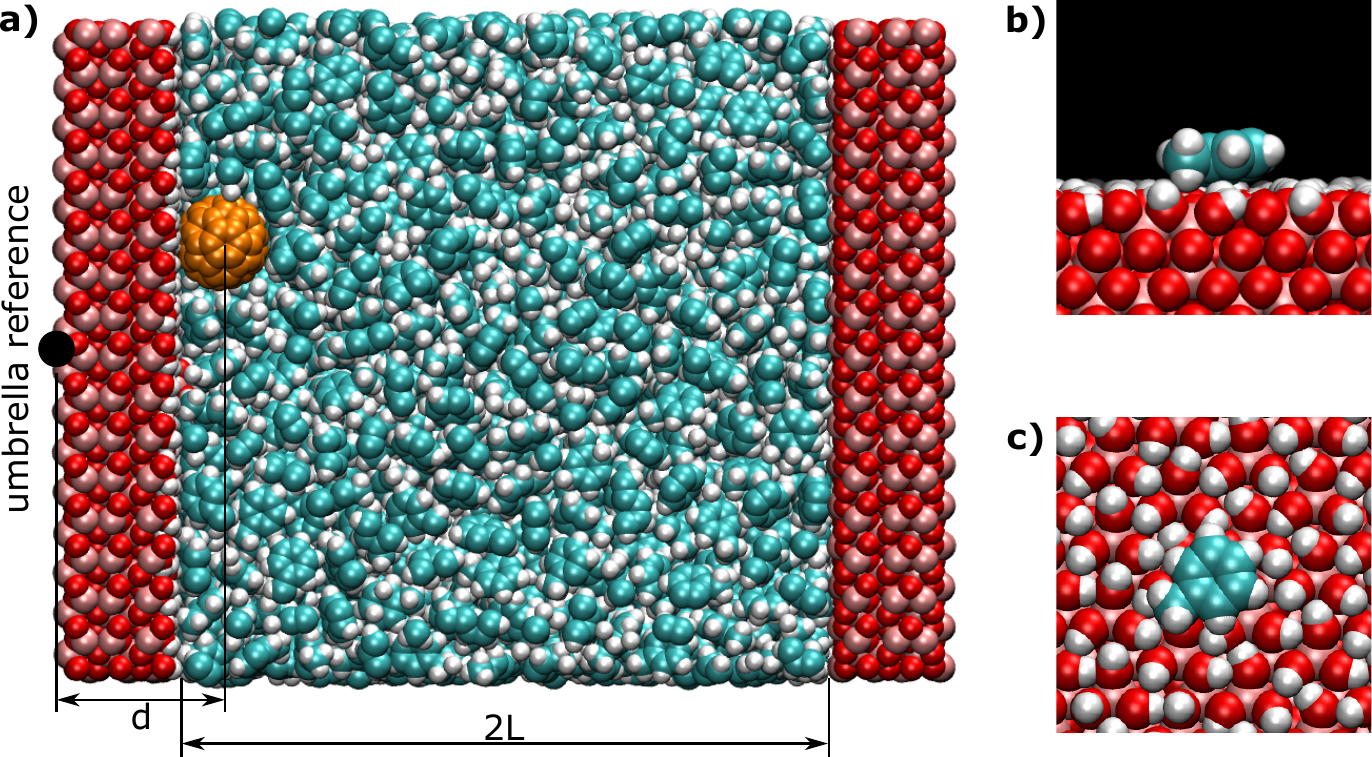}
\caption{\textbf{Representation of the simulated alumina slit pore.}
\textbf{a)} Sample system of the simulation containing one C60 fullerene, 2010 toluene molecules and an alumina bulk. The C60, coloured in orange, is close to the hydroxylated alumina surface. The width of the pore is indicated as well as the distance $d$ that is constrained with a harmonic potentail in the umbrella sampling simulations.
\textbf{b, c)} Orientation of an exemplary toluene molecule on the hydroxylated surface, with a side view (\textbf{b}) and top view (\textbf{c}).
}
\label{fig:pmf:system}
\end{figure}

We apply this approach to transport of fullerenes dissolved at finite dilution in toluene in hydroxylated alumina pores (Fig.~\ref{fig:pmf:system}a). Specifically, we study C60 and C70 which were found to have quite diverse translocation velocities across the porous media, as shown by chromatographic studies~\cite{Stalling1993, Gasper1995, Saito2004, Zarzycki2007, Suess2019}.
We assess the likelihood of dimer formation under confinement and compare it with the data about agglomeration in bulk solutions~\cite{Ruoff1993, Smith1996, Yin1996, Korobov1999, Herbst2005, Sawamura2007, Banerjee2013}.  Interestingly, our results show that, for the case under study, dimer formation occurs only in the proximity of the precipitation concentration and hence are in practice hindered as long as the suspension is stable. Accordingly, we show that the transport of fullerenes in toluene is controlled by their single-molecule diffusion coefficient. 

\section{Simulation methods}\label{sec:methods}

We perform \gls{md} simulations of a fullerene suspended within a slit nanopore filled with toluene using the GROMACS simulation software \cite{Gromacs1,Gromacs2,Gromacs3,Gromacs4,Gromacs5,Gromacs6,Gromacs7}.
For toluene, the \gls{opls} \cite{OPLSAA1} force field is used
\ab{and as input structure, a box of liquid toluene, equilibrated at \SI{298.15}{\K} by \citet{Caleman2012}, is used to fill the pore.}
For the fullerenes, C60 and C70, the atoms of the aromatic carbon of the \gls{opls} force field are used together with a structure from \glsentryshort{nmr} data \cite{Yannoni1991, C70source}, as has been proposed by Monticelli \cite{Monticelli2012}.
The pore wall consists of bulk alumina (\ce{Al2O3}) with one hydroxylated surface and one non-hydroxylated, aluminium terminated surface, whose force field is taken from previous works \cite{Vucemilovic2019, Vucemilovic2020}. 

All simulations are run with a time step of \SI{1}{\fs} and bonds involving hydrogen atoms were turned into constraints
using the LINCS \cite{LINCS} algorithm.
The short range non-bonded interactions are calculated, using a \gls{lj} potential  switched smoothly to zero between \num{0.9} and \SI{1.2}{\nm} \cite{Christen2005}.
Long range electrostatic interactions are calculated with the \glsentryshort{pme} technique \cite{PME1,PME2}.

\ab{The equilibration of our system first involved minimizing the energy of the system using steepest descent.
A short run in the NPT ensemble is conducted for \SI{5}{\ns} to obtain the correct density at a pressure of \SI{1}{\bar}.
Hereby, periodic boundary conditions are applied in all three dimensions and a semiisotropic coupling is used such that the coordinates are only altered in the direction normal to the pore wall \cite{Huang2008, Chinappi2010}.
The initial velocities are generated according to a Maxwell distribution of the desired temperature of \SI{293.15}{\K}.
The leap-frog algorithm is used to integrate the equations of motion, whereas temperature and pressure are controlled with a Berendsen thermostat and barostat \cite{Berendsen1984}, respectively.
The time constants are \SI{1.0}{\ps} and \SI{50.0}{\ps} for temperature and pressure, respectively, and the coupling occurs every ten steps.
Using built-in tools of GROMACS, we compute the liquid density distribution across the pore to check for bulk toluene density in the center after the equilibration.
In all simulations, the obtained density is within \SI{1}{\percent} of the experimental bulk toluene density of \SI{867.1}{\kg\per\meter\cubed} \cite{Daridon2018}, which is also the estimated statistical accuracy of our density assessment.}

\ab{A simulated annealing step is subsequently conducted at constant volume to facilitate proper equilibration of the surface layers of toluene near the wall.
A stochastic integrator with a temperature coupling time of \SI{0.1}{\ps} is used to heat the system up to \SI{700}{\K} over \SI{3}{\ns}, where it is kept for another \SI{3}{\ns}.
Afterwards it is cooled to the final \SI{293.15}{\K} over \SI{4}{\ns}, where it is kept for another \SI{2}{\ns}.
As the final density in the center of the pore resembles bulk toluene density, we immediately continue with a final equilibration step.
At this point, one could consider a short NPT run, but we found that such a step does not affect the density profile throughout the pore.
Therefore we proceed straight into performing a \SI{1}{\ns} long NVT simulation  using a leap-frog integrator and a time step of \SI{1}{\fs}.
The temperature is kept at \SI{293.15}{\K} with a \gls{NH} thermostat \cite{NoseHoover1, NoseHoover2} with coupling every ten steps and a coupling time of \SI{3.5}{\ps}.
The production run proceeds with the same settings for \SIrange{10}{40}{\ns}.
}

In all simulations with constant volume, the inner part of the solid crystal is frozen as the force field used cannot properly reproduce the vibrational spectrum of the crystal. The surface layer and the interfacial hydroxyl groups are allowed to rotate and vibrate.

For the calculation of the Potential of Mean Force (PMF), umbrella sampling simulations are conducted.
A simulation with the particle pulled slowly through the pore from one wall to the other yields the starting configurations for the individual umbrella runs. In the umbrella runs, the \gls{com} of the particle is then constrained by a harmonic potential with the equilibrium position determined from the starting configuration.
For each system, \numrange{36}{53} windows are simulated.
The equilibrium distances are separated by \SI{0.05}{\nm} with force constants of \SIrange{5000}{8000}{\kJ\per\mol\per\nm\squared} in the region close to the surface.
In the bulk region, the windows are separated by \SIrange{0.05}{0.1}{\nm} with force constants of \SIrange{1500}{2000}{\kJ\per\mol\per\nm\squared}.
\ab{From the simulations, the histograms of position distributions are obtained and converted to a PMF using the weighted histogram analysis method (WHAM) \cite{Kumar1992, Hub2010}}.
Further details on the umbrella simulations can be found in the SI section~1 and
SI-Table~1.

The simulations to calculate the PMF between two fullerenes are done in an analogue way.
Hereby, the distance between the \gls{com} of the fullerenes is constrained with a harmonic potential.
In the case of two fullerenes adhered to the wall, the distance constraint is restricted to the plane parallel to the wall, while both fullerenes are constrained to the first potential minimum close to the wall using another harmonic potential with force constant \SI{4000}{\kJ\per\mol\per\nm\squared}.

\ab{For all PMF calculations, the statistical uncertainty is estimated by splitting the trajectory into three parts and running the WHAM analysis. We obtain three additional PMFs with the same positional resolution as the full-length PMF. For each bin of the PMF, the statistical uncertainty is estimated as the root mean square deviation of those three PMFs from the original one, that is calculated from the full-length trajectory. These uncertainties will then be used to retrieve
uncertainties for all subsequently calculated results (see section 4.2 for details).}

\section{Residence probabilities of nanoparticles: the four state model}\label{sec:fourstatemodel:full}
\begin{figure}
    \includegraphics[width=\textwidth]{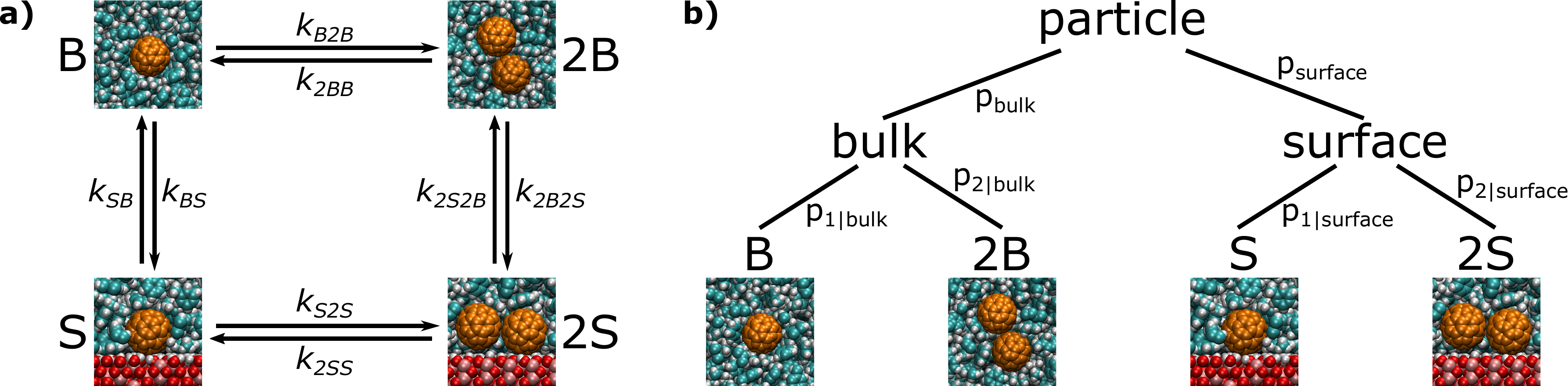}
\caption{\textbf{Graphical representation of the four state model.}
    \textbf{a)} The four state model with all non-zero transition rates.
    \textbf{b)} The probabilities accounted in the system.}
\label{fig:4-state-model}
\end{figure}
In order to model the residence probabilities of nanoparticles in the slit pore, we introduce discrete states covering the configuration space of the nanoparticles.
We restrict the model to the limit of low concentrations, where particle-particle interactions can be treated purely as pair-wise interactions.
Hence, we introduce the following four states as most occupied states for nanoparticles (cf. Fig.~\ref{fig:4-state-model}a):
\begin{itemize}
\item[B:] two fullerenes far away from each other and from the walls, i.e.  suspended in the ``bulk'' as monomers
\item[2B:] two fullerenes  bound to each other and suspended in the bulk.
\item[S:] two fullerenes trapped at the minimum of the fullerene-wall interaction and far away from each other. 
\item[2S:] two fullerenes bound to together and trapped at the minimum of the fullerene-wall interaction.
\end{itemize}
We then introduce transitions between each pair of states (cf. Fig.~\ref{fig:4-state-model}a).
The diagonal transitions correspond to two particles in the bulk state (i.e. separated by a large distance) directly forming a dimer located at the surface or vice versa.
This is assumed to be a composition of two processes, only occurring consecutively, not simultaneously.
Thus the diagonal transitions are considered very improbable and are neglected.

Using the transition rates $k_{ij}$ as defined in Fig.~\ref{fig:4-state-model}a, we can write the balance equations for the population probabilities of each state:
\begin{subequations}\label{eq:fourstatemodel_0}
\begin{align}\label{eq:fourstatemodel:full}
\dot{p}_B(t) &= - k_{BS}p_B(t)+k_{SB}p_S(t) -2k_{B2B}p^2_B(t)+2k_{2BB}p_{2B}(t)\\
\dot{p}_{2B}(t) &= +k_{B2B}p^2_B(t)-k_{2BB}p_{2B}(t)+k_{2S2B}p_{2S}(t)-k_{2B2S}p_{2B}(t)\\
\dot{p}_S(t) &= - k_{SB}p_S(t)+k_{BS}p_B(t)-2k_{S2S}p^2_S(t)+2k_{2SS} p_{2S}(t)\\
\dot{p}_{2S}(t) &= +k_{S2S} p_{S}^2(t) - k_{2SS}p_{2S}(t) +k_{2B2S}p_{2B}(t) - k_{2S2B}p_{2S}(t)
\end{align}
\end{subequations}
\ab{At equilibrium the particle concentrations are proportional to the Boltzmann weight $e^{-\beta U}$ where $U$ is the \gls{pmf}.
For such a concentration profile all local fluxes are vanishing.
This so-called detailed balance condition requires the fluxes in the two directions along each side to cancel each other.
Accordingly, for the first of the equations this implies}
\begin{equation}
    0 = \underbrace{ - k_{BS}p_B + k_{SB}p_S }_{=0} \: \underbrace{ -2k_{B2B}p^2_B + 2k_{2BB}p_{2B} }_{=0}
\end{equation}
and analogously for the other three equations.\\
Thus, eqs.~\eqref{eq:fourstatemodel_0}, can be rewritten as
\begin{subequations}\label{eq:fourstatemodel:pop}
\begin{align}
    p_B    &= \frac {k_{SB}   }{ k_{BS}   } p_S \label{eq:fourstatemodel:pop_pb}\\
    p_{2B} &= \frac{ k_{B2B}  }{ k_{2BB}  } p_B^2 \label{eq:fourstatemodel:pop_p2b}\\
    p_{2S} &= \frac{ k_{S2S}  }{ k_{2SS}  } p_S^2 \label{eq:fourstatemodel:pop_p2s}\\
    p_{2B} &= \frac{ k_{2S2B} }{ k_{2B2S} } p_{2S} \label{eq:fourstatemodel:pop_p2b_alt}
\end{align}
\end{subequations}
which are complemented by
\begin{equation}\label{eq:fourstatemodel:balance}
    1 = p_B + 2p_{2B} + p_S + 2p_{2S}\:.
\end{equation}
We can solve this set of equations for any of the populations, depending parametrically on the ratios of the rates.
Firstly, when inserting eqs.~\eqref{eq:fourstatemodel:pop_pb} to \eqref{eq:fourstatemodel:pop_p2s} into the balance equation~\eqref{eq:fourstatemodel:balance}, we obtain
\begin{equation}
    1 = \frac {k_{SB}   }{ k_{BS}   } p_S
      + 2 \frac{ k_{B2B}  }{ k_{2BB}  } \left( \frac {k_{SB}   }{ k_{BS}   } \right)^2 p_S^2
      + p_S
      + 2 \frac{ k_{S2S}  }{ k_{2SS}  } p_S^2\:,
\end{equation}
and thus
\begin{equation}
    2\left[ \frac{ k_{B2B}  }{ k_{2BB}  } \left( \frac {k_{SB}   }{ k_{BS}   } \right)^2
        + \frac{ k_{S2S}  }{ k_{2SS}  } \right] p_S^2
    + \left[ 1 + \frac {k_{SB}   }{ k_{BS}   } \right] p_S
    - 1
    = 0\:.
\end{equation}
\ab{While there are two solutions to this equation, one of which is strictly positive and the other negative.
As we require the probability $p_S$ to be positive, the solution is given as}
\begin{multline}\label{eq:fourstatemodel:solution}
    p_S =
    \frac{1}{4}
    \left[ \frac{ k_{B2B}  }{ k_{2BB}  } \left( \frac {k_{SB}   }{ k_{BS}   } \right)^2
            + \frac{ k_{S2S}  }{ k_{2SS}  } \right]^{-1}  \cdot
    \left\{-1 -\frac{k_{SB}}{k_{BS}}
    + \sqrt{ \left( 1 + \frac {k_{SB}   }{ k_{BS}   } \right)^2
            + 8 \left[ \frac{ k_{B2B}  }{ k_{2BB}  } \left( \frac {k_{SB}   }{ k_{BS}   } \right)^2
                     + \frac{ k_{S2S}  }{ k_{2SS}  } \right]} \right\}
\end{multline}
As the solution depends parametrically on the ratio of rates (e.g. $k_{SB}/k_{BS}$), which can be determined from ratios of probabilities according to eq.~\eqref{eq:fourstatemodel:pop}, the next step is to obtain expressions for these probabilities.
When picking a particle at random in the system, it can be either in a surface state or in a bulk state with corresponding probabilities $p_\textup{surface}$ and $p_\textup{bulk}$, respectively.
Additionally, it can be either an isolated particle or part of a dimer, with the probabilities $p_1$ and $p_2$, respectively (cf. Fig.~\ref{fig:4-state-model}b).
However, these can be different for the surface ($p_{1|\textup{surface}}$, $p_{2|\textup{surface}}$) and the bulk ($p_{1|\textup{bulk}}$, $p_{2|\textup{bulk}}$) states (see Fig.\ref{fig:4-state-model}).
We then obtain the probabilities for the four states by a combination of these as
\begin{subequations}\label{eq:fourstatemodel:concat-probabilities}
\begin{align}
    p_B     &= p_\textup{bulk}    \cdot p_{1}|_{\textup{bulk}}\:,\\
    2p_{2B} &= p_\textup{bulk}    \cdot p_{2}|_{\textup{bulk}}\:,\\
    p_S     &= p_\textup{surface} \cdot p_{1}|_{\textup{surface}}\:,\\
    2p_{2S} &= p_\textup{surface} \cdot p_{2}|_{\textup{surface}}\:.
\end{align}
\end{subequations}
At equilibrium, the probabilities $p_\textup{bulk}$ and $p_\textup{surface}$ as well as $p_1$ and $p_2$ will be given by the Boltzmann weights $\propto \exp(-\beta U)$, where $U$ is the associated \gls{pmf}.
Accordingly, the probability $p_\textup{bulk}$ is then given by 
\begin{equation}\label{eq:4sm:p-bulk}
    p_\textup{bulk}
    \propto \frac{ \int_{z^*}^{L} p(z) \d z }{ \int_{z^*}^{L} \d z }
    = \frac{1}{Z}\frac{  \int_{z^*}^{L}  \exp(-\beta U(z))  \d z }{ L - z^* }\:.
\end{equation}
Hereby, $Z=\int_{0}^{L} \exp(-\beta U(z))  \d z$ is the partition function, $L$ the half width of the pore, $z^*$ the parameter determining the boundary between the ``bulk'' and the ``surface'' state and $U(z)$ the PMF of a single particle with the surface (Fig.~\ref{fig:pmf:fullerenes-compare}a).
Naturally, the zero in $z$-direction is set to where the particle cannot further penetrate the wall.
Additionally, we obtain
\begin{equation}\label{eq:4sm:p-surf}
    p_\textup{surface}
    \propto \frac{1}{Z}\frac{  \int_{0}^{z^*} \exp(-\beta U(z)) \d z }{ z^* }\:.
\end{equation}
The (common) prefactor can be determined from the fact that each particle will be in one of the two states, i.e. $p_\textup{bulk} + p_\textup{surface} = 1$.
The ratio of the probabilities is given by
\begin{equation}
    K_{SB}
    := \frac{p_\textup{surface}}{p_\textup{bulk}}
    = \frac{ \int_{0}^{z^*} \exp(-\beta U(z)) \d z }{  \int_{z^*}^{L} \exp(-\beta U(z)) \d z } \cdot
      \frac{ L - z^* }{ z^* }\:,
\end{equation}
and
\begin{equation}
    p_\textup{bulk} = \frac{1}{1 + K_{SB}}\qquad , \qquad
    p_\textup{surface} = \frac{K_{SB}}{1 + K_{SB}}\:.
\end{equation}

Notably, here we recognize that $K_{SB}$ is the broadly known Henry's coefficeint in infinite dilution~\cite{Henry1803,Sander2015}, which is a measure for the attraction strength of the particle to the wall. This is of particular interest for chromatographic separation since it is a crucial parameter to determine the retention time across a chromatographic column~\cite{Nicoud2015,Reithinger2011,Michaud2021}

We continue with determining the probabilities $p_1$ and $p_2$ in an analogue fashion by introducing the dimer distance $R$ ($R_s$) below which two fullerenes are regarded as a dimer in bulk (surface). Accordingly we have: 
\begin{align}\label{eq:K21-bulk}
    K_{21}|_\textup{bulk}
    &:= \frac{ p_{2}|_{\textup{bulk}} }{ p_{1}|_{\textup{bulk}} }
    = \frac{ \int_0^{r^*} \exp (- \beta Y(r)) r^2\d r }
           { \int_{r^*}^R \exp (- \beta Y(r)) r^2\d r }
      \cdot \frac{R^3 - {r^*}^3 }{{r^*}^3}\:, \\
    K_{21}|_\textup{surface}
    &:= \frac{ p_{2}|_{\textup{surface}} }{ p_{1}|_{\textup{surface}} }
    = \frac{ \int_0^{r_S^*}     \exp (- \beta W(r)) r\d r }
           { \int_{r_S^*}^{R_S} \exp (- \beta W(r)) r\d r }
      \cdot \frac{R_S^2 - {r_S^*}^2 }{{r_S^*}^2}\:,\label{eq:K21-surface}
\end{align}
and
\begin{align}
    p_{1}|_\textup{bulk} &= \frac{1}{1 + K_{21}|_\textup{bulk}} , &
    p_{2}|_\textup{bulk} &= \frac{K_{21}|_\textup{bulk}}{1 + K_{21}|_\textup{bulk}}\:,\\
    p_{1}|_\textup{surface} &= \frac{1}{1 + K_{21}|_\textup{surface}} , &
    p_{2}|_\textup{surface} &= \frac{K_{21}|_\textup{surface}}{1 + K_{21}|_\textup{surface}}\:.
\end{align}
Hereby, $Y(r)$ is the PMF between two particles in the bulk liquid and $W(r)$ is the PMF between two particles when both are adhered to the surface.
The parameters $r^*$ and $r_S^*$ determine the boundary between the state of two adhered and two isolated particles for the bulk and at the surface, respectively.
These are analogous to $z^*$ and all three are taken to be the position of the edge of the interface layer. A natural choice is the maximum in the PMF. As it will be further discussed in the following section, for the case of fullerenes, the significant structuring occurs for two layers, and hence the maximum of the second potential barrier of the respective PMF is chosen to denote the interface region (see Fig.~\ref{fig:pmf:fullerenes-compare}).

In both of the latter cases, $R$ and $R_S$ 
are determined implicitly by the concentration.
The volume integral runs over the volume corresponding to a single particle, i.e. for a number concentration $c$, the integration boundaries are given by
\begin{align}
    R &= c^{-1/3}\:,\\
    R_S &= (c\cdot z^*)^{-1/2}\:.
\end{align}
We can express the ratios of transition rates in eq.~\eqref{eq:fourstatemodel:pop} via the ratios of probabilities of the four states:
\begin{align}
    \frac{ k_{SB} }{ k_{BS} }
    &= \frac{1}{ K_{SB} }
       \cdot \frac{ 1 + K_{21}|_\textup{surface} }{ 1 + K_{21}|_\textup{bulk} }\:,\\
    \frac{ k_{B2B} }{ k_{2BB} }
    &= \frac{1}{2} K_{21}|_\textup{bulk} \cdot ( 1 + K_{21}|_\textup{bulk} ) ( 1 + K_{SB} )\:,\\
    \frac{ k_{S2S} }{ k_{2SS} }
    &= \frac{1}{2} K_{21}|_\textup{surface} \cdot ( 1 + K_{21}|_\textup{surface} )
       \frac{ 1 + K_{SB} }{ K_{SB} }\:.
\end{align}
Here we notice that the quantities $K_{21}|_\textup{bulk} = 2p_{2B}/p_B$ and $K_{21}|_\textup{surface} = 2p_{2S}/p_S$ are the analogues to $K_{SB}$ -- they are a measure of the strength of agglomeration of the nanoparticles in the bulk and in the interface layer, respectively.\\

\section{Fullerenes in a toluene-filled alumina pore}
\subsection{Effective potentials}
%
\begin{figure}
\centering
\includegraphics[width=0.7\textwidth]{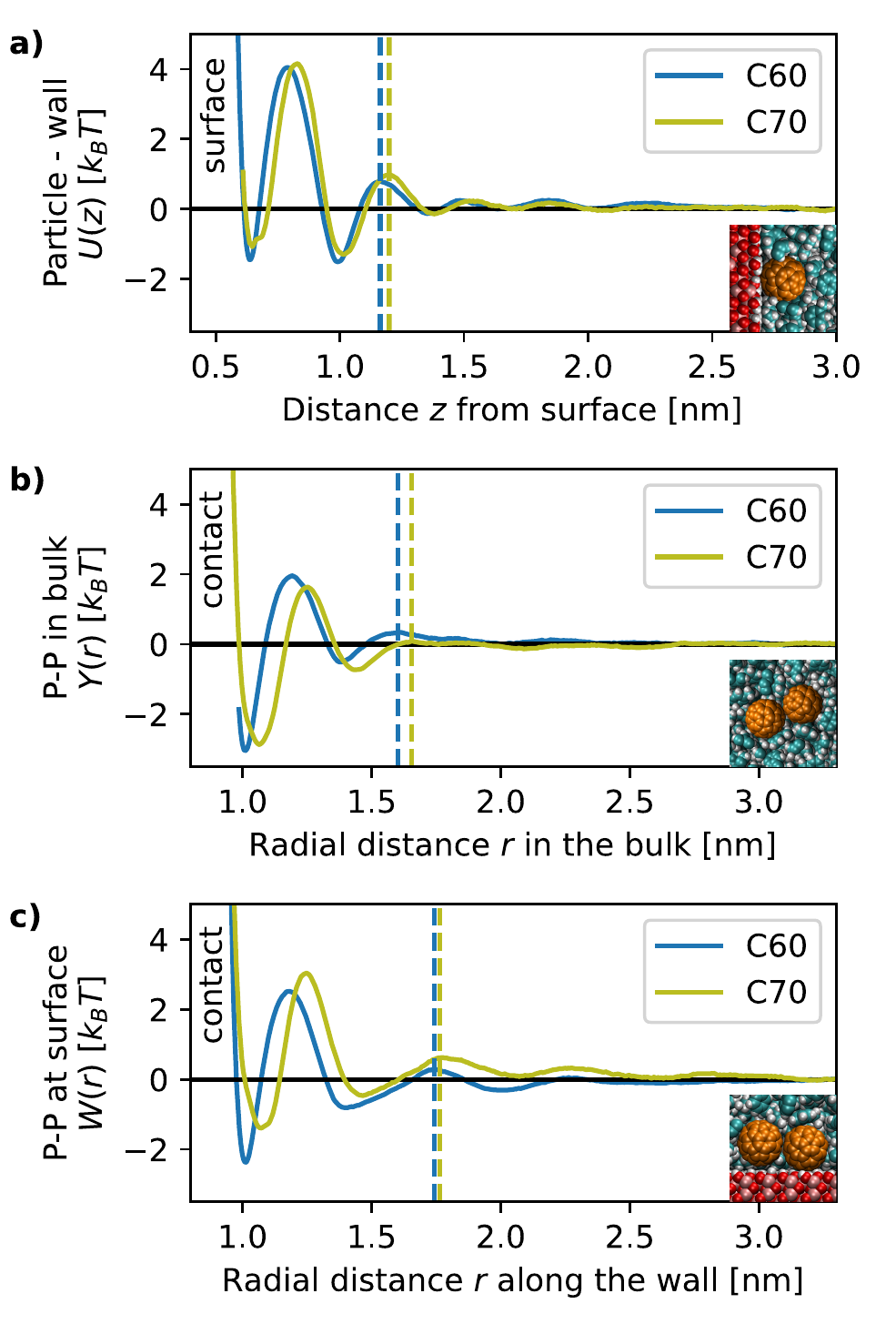}
\caption{\textbf{Characterisation of particle-pore and inter-particle interaction potentials for C60 and C70.}
The dashed lines indicate the position of separation between the two states ($z^*$, $r^*$ and $r_S^*$).
	\textbf{a)} Effective potential $U(z)$ between a nanoparticle and the hydroxylated alumina surface.
	\textbf{b)} Effective potential $Y(r)$ between two nanoparticles of the same kind in bulk liquid, as a function of their separation $r$.
	\textbf{c)} Effective potential $W(r)$ between two identical particles residing in the first minimum of $U(z)$, as a function of their lateral separation $r$.}
\label{fig:pmf:fullerenes-compare}
\end{figure}

We now proceed to apply the four state model to the transport of fullerenes through a pore. We first calculate the residence probabilities for C60 and C70 from the respective PMFs $U, Y, W$.
The latter are obtained from MD simulations with umbrella sampling for the two fullerenes C60 and C70 in bulk toluene and in a toluene filled slit nanopore with a hydroxylated sapphire alumina wall.

As shown in Fig.~\ref{fig:pmf:fullerenes-compare}, $U(z)$ possesses a minimum in the PMF at direct contact with the wall.
It is separated by a high barrier from a secondary global minimum for both C60 and C70, where the fullerene particle is separated from the wall by a solvent layer that is 1 toluene molecule thick, mostly adoptiong planar alignment of the aromatic carbon ring to the wall (Fig.~\ref{fig:pmf:system}b,c). The two minima are followed by smaller oscillations around the bulk value at larger distances from the wall.
This suggests that fullerenes will readily access the two surface states from the bulk, and hence they are both taken to be part of the interface layer. However, the depth of these minima is such that at room temperature the fullerenes will spend only limited time at the pore walls.  

\begin{figure}
\centering
\includegraphics[scale=1]{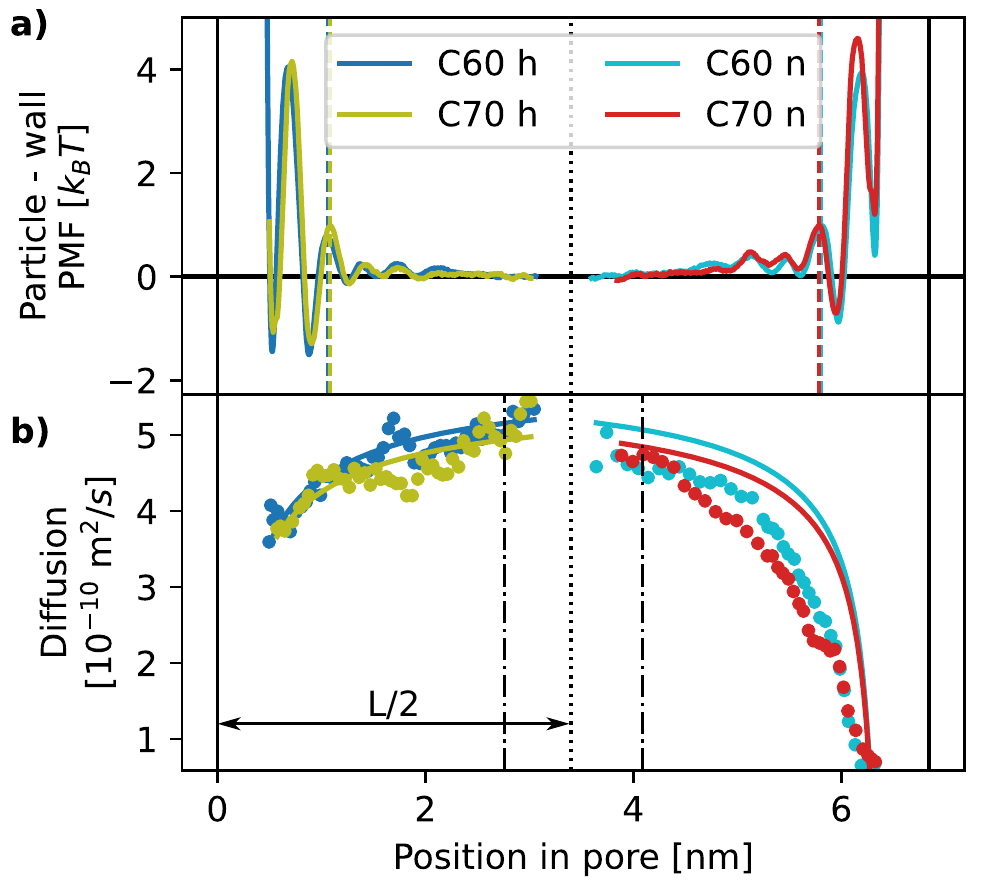}
\caption{
\textbf{Static and dynamic properties of a single fullerene in a slit pore with a hydroxylated (left) and aluminium-exposed wall (right).}
The vertical solid lines denote the edges of the pore, while the vertical dotted line denotes its center.
\textbf{a)} Potential of mean force (PMF) of a C60 and C70 througout the pore.
The dashed vertical lines indicate $z^*$, the separation between bulk and surface states.
\textbf{b)} The position dependent lateral diffusion coefficient, for both C60 and C70. The full lines are fit functions to the Faxén-Feitosa model (eq.~\ref{eq:feitosa}), while the symbols are data extracted from simulations.
Vertical dashed-dotted lines enclose the central region for calculation of $D_\textup{centre}$.
}
\label{fig:pore:C60-C70}
\end{figure}

The PMFs of C60 and C70 differ by three main aspects.
First, the larger size of C70 shifts the PMF away to slightly larger separations as compared to C60.
Second, the anisotropy of C70 induces a small inhomogeneity in the first minimum, relating to the different possible orientational configurations (Fig.~\ref{fig:pore:C60-C70}), with the two principle axes parallel to the walls.
Third, the depth of the minima of C70 are smaller and the height of the barriers are larger as compared to C60. 

We then proceed to calculate the PMF between two C60 in bulk liquid (Fig.~\ref{fig:pmf:fullerenes-compare}b) to obtain an estimate for the enthalpy of solvation.
We find a value of $\Delta E = \SI{-7.4+-0.6}{\kJ\per\mol}$\ab{(note: included error estimate)}, which is in good agreement with the enthalpy of solvation of C60 in toluene determined by experiments to be about \SI{-8.7}{\kJ\per\mol} \cite{Ruoff1993, Smith1996, Yin1996, Korobov1999, Herbst2005, Sawamura2007} (see SI section~2 for details).
We note that Banerjee \cite{Banerjee2013} reported PMF data with lower attractive strength of only about \SI{4}{\kJ\per\mol}.
While the procedure to calculate the PMF is similar to our case, the force field used differs, which probably causes the differences.

In the following we complement the PMF of two nanoparticles in toluene in the bulk ($Y(r)$) with the PMF between two nanoparticles at the interface ($W(r)$).
Like in $U(z)$, we find two well defined minima in both $Y(r)$ and $W(r)$, except that now the two particles in contact make up a dominant state, while the separation by one toluene molecule is higher in free energy (cf. Fig.~\ref{fig:pmf:fullerenes-compare}).
In the bulk we find that the height of the barriers between these two states of C70 is slightly smaller than these between two C60, a feature that is opposite at the interface.
The depth of these minima, however, suggests that at room temperature, dimers will be only transient.

\subsection{The application of the four state model}
With these PMFs, we can resort to the four state model for both C60 and C70, to calculate the Henry's coefficient and estimate the likelihood for the agglomeration of fullerenes in bulk and at the interfaces.  We present the results for  C60 and C70 at concentrations of \SI{0.014}{\kg\per\m\cubed} (Table~\ref{tab:solution-4sm}), which is about \SI{1}{\percent} of the solubility limit of C70 \cite{Zhou1997}. As a result \ab{(upper part of Table}~\ref{tab:solution-4sm}\ab{)}, we obtain the fraction of fullerene particles in each of the four states: a single particle in the bulk, on the surface, or as dimers in the bulk or on the surface, as stated by eq.~\eqref{eq:fourstatemodel:concat-probabilities}.
From this data, at the precipitation limit, the fraction of dimers seems to be significant for both particle types.
However, away from the limit the fraction of dimers drops significantly (see SI section~3 and SI-Fig.~1 for details).
\ab{Hereby, we estimate the confidence of all calculated values by disturbing all involved PMFs up to their estimated uncertainty (cf. section~\ref{sec:methods}), one at a time, and solving the four state model.
The confidence interval is determined from the root mean square deviation of all perturbed results from the unperturbed result.}

Further analysis of the data allows us to report the Henry's coefficients \ab{(lower part, column 1 of Table}~\ref{tab:solution-4sm}\ab{)} $K_{SB} = p_\textup{surface} / p_\textup{bulk}$ for C60 and C70 in these pores.
As expected, $K_{SB}$ are very similar for the two particles, differing by less than \SI{10}{\percent} \ab{and fully within the statistical uncertainty}.
Thus we do not expect large differences in their adsorption behaviour.
Consistently, the associated retention times 
adopt the same value for C60 and C70 for the present data (\ab{lower part,} last column in Table~\ref{tab:solution-4sm}).
Therefore, we do not expect chromatographic separation to happen in this type of setup, which is indeed the case.
Actually, more complex surface functionalizations, mostly relying on polyaromatic hydrocarbons or long carbon chains are used for the separation of C60 from C70 \cite{Stalling1993, Gasper1995, Saito2004, Suess2019}.

To measure the strength of agglomeration in the bulk and at the pore wall, we calculate the quantities $K_{21}|_\textup{bulk}$ and $K_{21}|_\textup{surface}$, respectively, following eqs.~\eqref{eq:K21-bulk} and \eqref{eq:K21-surface}.
We notice that $K_{21}|_\textup{bulk}$ is larger than one for C60 and C70, thus the particles favor the adhered state.
When comparing the values for C60 and C70 (\ab{lower part,} column \ab{3} in Table~\ref{tab:solution-4sm}), we see, that the agglomeration of C70 in bulk toluene is stronger than this of C60, which is also consistent with the smaller solubility of C70 than of C60 \cite {Zhou1997}.

In contrast, when looking at fullerenes adhered to a surface, a significant drop in the $K_{21}$ coefficient is observed compared to bulk values (\ab{lower part,} column \ab{4} in Table~\ref{tab:solution-4sm}).
Actually, for C70 $K_{21}|_\textup{surface}$ is smaller than one, suggesting that particles favor the monomeric state, as a consequence of the more shallow minimum in the PMF.
Thus, we would expect an overall dissociation of the fullerenes in small pores, where the surface state dominates the bulk state.
In any case the difference of the weight of the dimers compared to that of the monomers $p_\textup{1}\cdot V_\textup{1}$ is negligible as it is less than \SI{2}{\percent} for both particles and up to the solubility limit, whereas it increases beyond (see SI section~3 and SI-Fig.~1 for details).

\begin{table}
\centering
\begin{tabular}{l||
                SSSS}
 &
{$p_B$} &
{$p_S$} &
{$2p_{2B}$} &
{$2p_{2S}$} \\
Particle &
[\si{\percent}] &
[\si{\percent}] &
[\si{\percent}] &
[\si{\percent}]
\\
\midrule
C60 &
19.6+-1.7 & 27.3+-0.8 & 24.4+-1.7 & 28.7+-0.8
\\
C70 &
16.3+-1.5 & 31.2+-3.1 & 30.0+-1.9 & 22.5+-2.9
\end{tabular}

\begin{tabular}{l||
                SSS|
                S}
 &
&
\multicolumn{2}{c|}{$K_{21}$}
& \\
Particle &
{$K_{SB}$} &
{bulk} &
{surface} &
{$t_i / t_0$}
\\
\midrule
C60 &
1.28+-0.07 & 1.24+-0.35 & 1.05+-0.08 & 1.26+-0.02
\\
C70 &
1.16+-0.16 & 1.84+-0.38 & 0.72+-0.28 & 1.26+-0.12
\end{tabular}
\caption{\ab{Numbers are updated to include error estimates. All references to the table are updated to the two-part layout.}
Solution of the four state model for the probabilities of a nanoparticle to be in one of the four states.
The number concentration is \SIlist{1.17e-5;1.00e-5}{\per\nm\cubed} for C60 and C70, respectively, resulting in a concentration of \SI{0.014}{\kg\per\m\cubed}.
The width of the pore is \SI{6.8}{\nm}.
}
\label{tab:solution-4sm}
\end{table}


\section{Transport  of fullerenes through slit nanopores}\label{sec:pore:results}
Our final aim is to utilize the four state model to predict the effective diffusion constant of the nanoparticle in the pore.
We focus on transport parallel to the pore walls, where the diffusion coefficient can be extracted directly from the umbrella sampling simulations (see methods section~2 and SI section~1).
To simplify the analysis, we focus on concentrations below the precipitation limit, such that the fullerenes are mainly in the single molecule state. Consequently, contributions of dimers to the overall transport can be omitted. 

In order to asses the role of the potential and the surface functionalisation, we complement our hydroxylated alumina wall with a wall where the alumina surface is terminated by aluminium atoms carrying partial charge (with the system being overall neutral). The overall width of the pore is \SI{6.8}{\nm} to ensure bulk properties of toluene in the middle of the pore and no cross-talk of the two solid-liquid interfaces (Fig.~\ref{fig:pore:C60-C70}a). Similar to the hydroxylated wall, the wall with  exposed aluminium also demonstrates two minima separated by a high barrier. However, for both C60 and C70 the minimum where the fullerenes are in contact with the wall is unstable relative to the bulk. Besides the second shallow minimum, the potential is overall repulsive, for C70 more so than for C60.

\subsection{Diffusion profiles of nanoparticles in pores}\label{sec:pore:profiles}
We now use our umbrella sampling to extract from the simulations the diffusion profiles $D(z)$ of fullerenes in the pore as a function of the the distance of their center of mass from the wall (symbols in Fig.~\ref{fig:pore:C60-C70}b).
While providing interesting information about the interplay between effective interactions and diffusivity, this data will also serve as a benchmark in the comparison with our modeling efforts.
\begin{table}
\centering
\begin{tabular}{l|SS|S}
Particle / Surface &
{$D_\textup{centre}$} &
{$D_\textup{srfc}$} &
{$z_0$}\\ &
\multicolumn{2}{c|}{[\SI{e-10}{\m\squared\per\s}]} &
{[\si{\nm}]} \\
\midrule
C60 / hydroxylated & 5.0 & 4.2 & 0.26+-0.02 \\
C70 / hydroxylated & 4.9 & 3.8 & 0.31+-0.04 \\
C60 / aluminium exposed & 5.0 & 0.7 & -0.18+-0.01 \\
C70 / aluminium exposed & 4.9 & 0.7 & -0.18+-0.01 \\
\end{tabular}
\caption{
Characteristics of the diffusion profiles of C60 and C70 at the hydroxylated and aluminium exposed wall. For reference $D_\textup{bulk}^{\textup{C}60} = \SI{5.7e-10}{\m\squared\per\s}$, while $D_\textup{bulk}^{\textup{C}70} = \SI{5.5e-10}{\m\squared\per\s}$.
The diffusivities at the centre of the pore $D_\textup{centre}$ and at the pore surface $D_\textup{srfc}$ are extracted from simulations, while the slip length $z_0$ is obtained from fitting eq.~\eqref{eq:feitosa} to the data.
\ab{The error estimates on the slip length are solely the uncertainty of fitting parameter, errors inherited from the diffusion coefficient or the definition of the pore wall are not taken into account.}}
\label{tab:average-diffusion}
\end{table}

First, we analyze the diffusivity of C60 and C70 in the pore centre (denoted as $D_\textup{centre}$), where the effective potentials of both walls vanish (column 2 in Table~\ref{tab:average-diffusion}).
In the range of positions in the pore of \SI{2.8}{\nm} until \SI{4.2}{\nm} we find $D_\textup{centre}^\textup{C60} = \SI{5.0e-10}{\m\squared\per\s}$ and $D_\textup{centre}^\textup{C70} = \SI{4.9e-10}{\m\squared\per\s}$.
This compares reasonably well to bulk diffusion coefficients
$D_\textup{bulk}^\textup{C60} = \SI{5.7e-10}{\m\squared\per\s}$, $D_\textup{bulk}^\textup{C70} = \SI{5.5e-10}{\m\squared\per\s}$ obtained from independent simulations of a fullerene in simple bulk liquid.

The bulk diffusion coefficients are themselves in reasonable agreement \cite{Pearson2018} with the prediction from the Stokes-Einstein-Sutherland formula $D_\textup{SES} = \kbT / 6\pi\eta R$.
Here, the viscosity $\eta$ was determined using the Green-Kubo formula to be \SI[inter-unit-product=\cdot]{8.5e-4}{\Pa\s}, while the hydrodynamic radii $R$ of C60 and C70 are set to \SIlist{0.52; 0.54}{\nm}, respectively (see SI section~4 for details).
Consequently, $D_\textup{SES}^\textup{C60} = \SI{4.9e-10}{\m\squared\per\s}$ and $D_\textup{SES}^\textup{C70} = \SI{4.7e-10}{\m\squared\per\s}$.

Notably, there is a small but significant difference of the order of \SI{10}{\percent} between $D_\textup{bulk}$ and $D_\textup{centre}$ for both C60 and C70.
This shows that, while from the structural point of view, the toluene solvent is indeed in the bulk state, the effects of the pore walls on the diffusive transport are significantly more long ranged.

We now focus on the diffusivity at the pore wall $D_\textup{srfc}$ (column 3 in Table~\ref{tab:average-diffusion}).  Here, we set the zero distance from the surface ($z=0$) by the position of the outer edge of the fullerene located in the first minimum of the PMF. This provides a good reference state for all systems considered. With this definition, we find that the diffusion coefficient of both fullerene types in close proximity to the wall drops due to the surface friction, as expected. Interestingly, however, the decrease is  less than \SI{30}{\percent} at the hydroxylated surface, where the diffusivity adopts a value of $D_\textup{srfc} = \SI{4.2e-10}{\m\squared\per\s}$ and $D_\textup{srfc} = \SI{3.8e-10}{\m\squared\per\s}$ for C60 and C70, respectively (see Fig.~\ref{fig:pore:C60-C70}b).
On the other hand, on the surface with exposed charges, $D_\textup{srfc} = \SI{0.7e-10}{\m\squared\per\s}$ for both particles, which is a drop by about one order of magnitude compared to $D_\textup{centre}$.
\ab{This suggests that the toluene-wall slip length is substantially positive} on the hydroxylated wall and \ab{close to zero} at the wall exposing the partially charged aluminium atoms.

Between the surface and the centre of the pore, the diffusion profiles $D(z)$ monotonously increase (within the statistical accuracy of the extracted data). On the hydroxylated wall, the C60 and the C70 profile are basically indistinguishable, while at the aluminium exposed wall, C70 is clearly somewhat slower than C60.  

The above described key features of the profiles can be captured 
by a so called Faxén-Feitosa model that describes the change of the parallel diffusion coefficient of a colloidal sphere near a flat surface \cite{Faxen1923,Feitosa1991}.
Accordingly,
\begin{equation}\label{eq:feitosa}
D(z) = D_\textup{bulk}\; \zeta_\parallel (z)\:,
\end{equation}
where
\begin{align*}
\zeta_\parallel (z)
= \left[ 1 - \frac{9}{16} \left(\frac{R}{z - z_0}\right)
+ \frac{1}{8} \left(\frac{R}{z - z_0}\right)^3
- \frac{45}{256} \left(\frac{R}{z - z_0}\right)^4
- \frac{1}{16} \left(\frac{R}{z - z_0}\right)^5 \right]\:.
\end{align*}
Hereby, $z_0$ is the position at which the no-slip condition is imposed, which is the slip length if the wall is at $z=0$.
Naturally, in the case of a vanishing slip length, $z_0=0$.

Using the previously reported hydrodynamic radii and the bulk diffusion constants, eq.~\eqref{eq:feitosa} can be applied to data with $z_0$ as a single fitting parameter. 
For the hydroxylated side, the model agrees well with the data (full lines in Fig.~\ref{fig:pore:C60-C70}\ab{b}) and we find a slip length $z_0 = \SI{0.26+-0.02}{\nm}$ and \SI{0.31+-0.04}{\nm} for C60 and C70, respectively (column 4 in Table~\ref{tab:average-diffusion}), which is comparable to the thickness of a layer of toluene.
\ab{A small but positive slip length is indeed expected from the relation of the slip length to the contact angle \cite{Huang2008} and measurements of the contact angle of toluene on the similar surface of anodized aluminum oxide that yield contact angles on the order of \SI{60}{\degree} to \SI{70}{\degree} \cite{Redon2005}.}

In contrast, on the non-hydroxylated side, the agreement of the model is worse. The obtained slip length is slightly negative ($z_0= \SI{-0.18+-0.01}{\nm}$), \ab{with the error estimate reflecting only the uncertainty of the fitting procedure.
While this result might be an artifact due to the model not reproducing the data particularly well, negative slip lengths were already obtained before in systems with highly wettable surfaces \cite{Huang2008, Chinappi2010, Kong2010}.
Nonetheless, in these studies, the obtained (negative) slip lengths were also smaller but of comparable size as the solvent molecules.
To further investigate the presence of a negative slip length, a set of different simulations can be conducted, advantageously taking into account the relation between the slip length and the contact angle \cite{Huang2008}.
However, these studies are kept for future work.}

\ab{Despite the discrepencay on the non-hydroxylated side}, the Faxén-Feitosa model well captures the slow relaxation of the diffusion constant within the pore from the \ab{hydroxylated} surface value to the bulk value. Interestingly however, inspection of Fig.~\ref{fig:pore:C60-C70}b shows that the range at which the diffusivity is affected by the presence of \ab{each} wall is very weakly dependent of the slip length at the wall. Further inspection of the diffusivity of the solvent shows that, actually, the effect of slow relaxation can be associated with the viscosity gradient in the toluene (see SI section 5 and SI-Fig.~2 for details).
As such this is directly a consequence of the lack of length scale separation in the system.

\ab{In order to further tell apart the effects of molecular interaction on the alteration of the viscosity from those of the boundary conditions, one could combine the present approach with different multiscale approaches like Langevin dynamics.
As in such simulations, the variation of the viscosity can be explicitly taken into account or be ignored, different effects altering the diffusion coefficient can be told apart \cite{Gubbiotti2019}.
Nonetheless, a technique capable of retrieving the \gls{pmf} is required to apply the four state model and to calculate the diffusion coefficients averaged over the whole pore as discussed in the following section.}

\subsection{Effective diffusion coefficients}\label{sec:pore:averages}
\begin{table}
\centering
\begin{tabular}{l|S
                  S[table-format=1.1, table-number-alignment = left]
                  S[table-format=1.1, table-number-alignment = left]}
Particle / Surface &
{$\langle D(z) \rangle$} &
{$D_\textup{eff}$ (err$\%$)} &
{$D_\textup{app}$ (err$\%$)} \\
\midrule
C60 / hydroxylated & 4.8 & 4.8 $\;(<1\%)$ & 4.5 $\;(6\%)$ \\
C70 / hydroxylated & 4.7 & 4.7 $\;(<1\%)$ & 4.4 $\;(6\%)$ \\
C60 / aluminium exposed & 4.1 & 4.5 $\;(10\%)$ & 5.0 $\;(23\%)$\\
C70 / aluminium exposed & 4.0 & 4.5 $\;(10\%)$ & 4.9 $\;(21\%)$
\end{tabular}
\caption{Average diffusion coefficients. All values are in units of \SI{e-10}{\m\squared\per\s}. The errors are given as the relative deviation from the reference $\langle D(z)\rangle$.
\ab{The values of both the effective and the apparent diffusion coefficient depend on the choice of the parameters $z^*$, $r^*$ and $r$ (cf. section~\ref{sec:fourstatemodel:full}).
To explore this effect, we explicitly perturb their values within their confidence intervals (i.e. the uncertainty of the position of the respective maxima in the PMF).
Hereby, the values for the diffusion coefficients $D_\textup{eff}$ and $D_\textup{app}$ always stay with \SI{1}{\percent} of the presented values.}
}
\label{tab:averages}
\end{table}

We finally proceed with calculating the average diffusion coefficient of a nanoparticle in the pore $\langle D \rangle$.
The latter can be obtained directly from the diffusion profile $D(z)$, weighted with the probability to find a particle at the given distance from the wall, which is captured by  $U(z)$
\begin{equation}\label{eq:diff-weighted-average}
    \langle D \rangle = \frac{\int_0^L D(z) \exp(-\beta U(z)) \d z}{\int_0^L \exp(-\beta U(z)) \d z}\:.
\end{equation}
Here, $\beta$ is the inverse of the thermal energy $\kbt$ and the integration is performed to the centre of the pore. 
The mean diffusion values are included in Table~\ref{tab:averages} (column 2) and clearly demonstrate the strong difference between the hydroxylated side and the non-hydroxylated side of the pore.
While for the hydroxylated side, the mean diffusion is about \SI{15}{\percent} smaller than the bulk value, for the non-hydroxylated side, it is reduced by about \SI{30}{\percent}.
While most accurate, obtaining these values is associated with a considerable computational effort.
It is therefore useful to consider protocols that provide good approximates.  

One possible approximation can be established by the use of the four state model (cf. section~\ref{sec:fourstatemodel:full}). In the absence of dimers, this model reduces to two states -- the nanoparticle being in the bulk region of the pore or in the interface region, with probabilities  $p_\textup{bulk}$ and $p_\textup{surface}$, as defined by eqs.~\eqref{eq:4sm:p-bulk} and \eqref{eq:4sm:p-surf}.
Rather than integrating $D(z)$, the diffusivity in these two compartments is represented by 
$D_\textup{centre}$ and $D_\textup{srfc}$. Hence, the effective diffusion coefficient can be simply obtained by
\begin{equation}\label{eq:diff-effective}
    D_\textup{eff}
    = \frac{p_\textup{surface}\cdot V_\textup{surface}\cdot D_\textup{srfc} + 
            p_\textup{bulk}\cdot V_\textup{bulk}\cdot D_\textup{centre}}
           {p_\textup{surface}\cdot V_\textup{surface} + 
            p_\textup{bulk}\cdot V_\textup{bulk}}\:.
\end{equation}

For the hydroxylated pore,  $D_\textup{eff}$ gives a very accurate prediction of the average diffusion coefficient $\langle D(z) \rangle$ differing by less than \SI{1}{\percent}.
For the aluminium exposed pore, $D_\textup{eff}$ overestimates the average diffusion coefficient by about \SI{10}{\percent}.
These overestimates come due to the neglect of the decreasing diffusion coefficient between the surface state and the center of the pore, which is more prominent for the system where there is basically no slip on the surface.

An alternative approximation for $\langle D(z) \rangle$, commonly used in separation technology, involves setting $D_\textup{srfc}$ to zero.
Accordingly, the nanoparticle can either be mobile in the center of the pore or adhered at the wall that inhibits all motion.
The apparent transport coefficient $D_\textup{app}$ is then given as the diffusivity of the particle in the bulk liquid reduced by the fraction of time spent on the surface
\begin{equation}\label{eq:diff-apparent}
    D_\textup{app}
    = \frac{p_\textup{bulk}\cdot V_\textup{bulk}\cdot D_\textup{bulk}}
           {p_\textup{surface}\cdot V_\textup{surface} + 
            p_\textup{bulk}\cdot V_\textup{bulk}}=
            \frac{D_\textup{bulk}}
           {K_\textup{SB}\frac{V_\textup{surface}}{V_\textup{bulk}}+1}\:.
\end{equation}
On the right hand side, $D_\textup{app}$ is simply expressed through the Henry's coefficient. Naturally, this approximation is expected to work very well in thick pores, and where there is no-slip at the wall, i.e. when there is a proper separation of length scales in the system.
However, in \ab{(slit)} nanopores, the validity of this approximation is not guaranteed.

In the pore of intermediate thickness and partial slip, typical for nanoparticles, $D_\textup{app}$ clearly underestimates the reference values $\langle D(z) \rangle$ (see column 2 and 4 in Table~\ref{tab:averages}), while it overestimates the reference in cases with no slip on the surface.
Actually, $D_\textup{app}$ is significantly outperformed by $D_\textup{eff}$, particularly on hydroxylated pore, due to substantial slip.

\section{Conclusions}
In the present work, we have addressed the problem calculating the average diffusion coefficient for transport of a nanoparticle along a pore. In  particular, we focused on the regime when there is no clear length scale separation between the particle size and the range of the potential, while the pore is considered sufficiently large such that bulk liquid is recovered in the central part of the pore, at least from the structural point of view.

We engage in two strategies - calculating the full diffusion profile using advanced sampling techniques in molecular dynamics simulations. We complement these simulations with a simplified model where the interaction and the diffusivity is coarse-grained into 4 states for which the characteristic parameters are extracted from MDs.
This strategy significantly reduces the computational effort compared to the reference, brute-force approach, yet provides a fully controlled approximation of the effective diffusion constant.
\ab{This is the case as sampling the diffusion coefficient to the desired degree of accuracy required at least ten times the simulation time needed for converging the PMF.
The use of equation~\eqref{eq:diff-effective} allows sampling the diffusion coefficient only in the center of the pore and at the wall, while in the intermediate positions only sampling the PMF.
For the system in study, using equation~\eqref{eq:diff-effective} needs only about one third of the simulation time compared to evaluating the full diffusion profile with equation~\eqref{eq:diff-weighted-average}.
}

As part of the four state model, we recover the Henry's coefficient and are able to provide an estimate for the particle retention time. Furthermore, it also allows us to account for the finite concentrations of nanoparticles. 
Despite its approximate nature, the model is still able to predict fine differences in the behaviour of similar nanoparticles, as exemplified on interactions and diffusivities of C60 and C70 fullerenes. 

Notably, the performance of the model depends on the friction properties of the pore surface, and not that much on the strength of the particle-wall potential. Specifically, if there is significant slip of the nanoparticle on the pore wall, the model performs particularly well. If the diffusivity at the interface drops significantly, the diffusivity tends to be overestimated. In all situations, however, the model outperforms some commonly used approximations.

This approach demonstrates that relying on a limited computational effort it is possible to recover the dynamic behaviour of nanoparticles in confinement.

\section{Acknowledgements}
We thank M. Sega and  K. Höllring for useful discussions. We are furthermore grateful to M. Han\v{z}eva\v{c}ki, R. Stepi\'{c} and N. Vu\v{c}emilovi\'{c}-Alagi\'{c} for preliminary calculations of effective potentials.
We acknowledge funding by the Deutsche Forschungsgemeinschaft (DFG, German Research Foundation) – Project-ID 416229255 – SFB 1411 Particle Design, (subprojects D1, D2 and B5).
We gratefully acknowledge the scientific support and HPC resources provided by the Erlangen National High Performance Computing Center (NHR@FAU) of the Friedrich-Alexander-Universität Erlangen-Nürnberg. 

\bibliography{bibliography.bib}

\appendix

\section{Additional information on the methods}
For the calculation of the \gls{pmf}, umbrella sampling simulations are conducted.
A simulation with the particle pulled slowly through the pore from one wall to the other yields the starting configurations for the individual umbrella runs.
The \gls{com} of the pore wall is used as a reference position.
In the umbrella runs, the \gls{com} of the particle is then constrained with a harmonic potential with the equilibrium distance to that reference determined from the starting configuration.
Hereby, only the direction \ab{normal} to the pore wall is taken \ab{for calculating the distance $d$}.
For each system, \numrange{36}{53} windows, with equilibrium distances separated by \SIlist{0.05; 0.1}{\nm} each, are simulated.
The force constants are \SIlist{5000;8000}{\kJ\per\mol\per\nm\squared} for the short distances and \SIlist{1500;2000}{\kJ\per\mol\per\nm\squared} for larger separations.
Further details on the simulated windows and the corresponding force constants can be found in SI-Table~\ref{tab:umbrella:windows}.
Each simulation is run for \SIrange{10}{40}{\ns} and the distance between particle and reference group is written every \num{100} steps (i.e. every \SI{100}{\fs}), while the positions of the whole system are only written every \SI{5}{\ps}.
The latter trajectory is then used to calculate the diffusion coefficient from the \gls{msd} of the particle.
\ab{Averaging over three neighboring windows was hereby necessary to converge the diffusion coefficient.}
\begin{table}
\centering
\begin{tabular}{l|ccc|ccc}
 & \multicolumn{3}{c|}{short separations} & \multicolumn{3}{c}{large separations} \\
System & no. of & force c. & sep. & no. of & force c. & sep. \\
 & wind. & \si[per-mode=fraction]{\kJ\per\mol\per\nm\squared} & [\si{\nm}] & wind. & \si[per-mode=fraction]{\kJ\per\mol\per\nm\squared} & [\si{\nm}] \\
\midrule
C60 hydroxy.     & 18 & 8000 & 0.05 & 35 & 2000 & 0.05 \\
C60 non-hydroxy. & 22 & 5000 & 0.05 & 17 & 1500 & 0.10 \\
C70 hydroxy.     & 18 & 8000 & 0.05 & 35 & 2000 & 0.05 \\
C70 non-hydroxy. & 22 & 5000 & 0.05 & 14 & 1500 & 0.10
\end{tabular}
\caption{Overview of the umbrella windows for the different systems.}
\label{tab:umbrella:windows}
\end{table}

\section{Enthalpy of solavtion of C60 in toluene}
Several experimental studies performed measurements of the enthalpy of solvation of C60 in toluene.
The values are
\SI{-11}{\kJ\per\mol} \cite{Ruoff1993},
\SI{-8.5 \pm 0.7}{\kJ\per\mol} \cite{Korobov1999},
\SI{-8.6 \pm 0.7}{\kJ\per\mol} \cite{Smith1996},
\SI{-7.62 \pm 0.01}{\kJ\per\mol} \cite{Yin1996},
\SI{-8.5 \pm 0.3}{\kJ\per\mol} \cite{Herbst2005},
\SI{-8 \pm 1}{\kJ\per\mol} \cite{Sawamura2007},
resulting in an average value of \SI{-8.7}{\kJ\per\mol}.

\section{Concentration dependence of the relative populations}
\begin{figure}
    \centering
    \includegraphics[width=\textwidth]{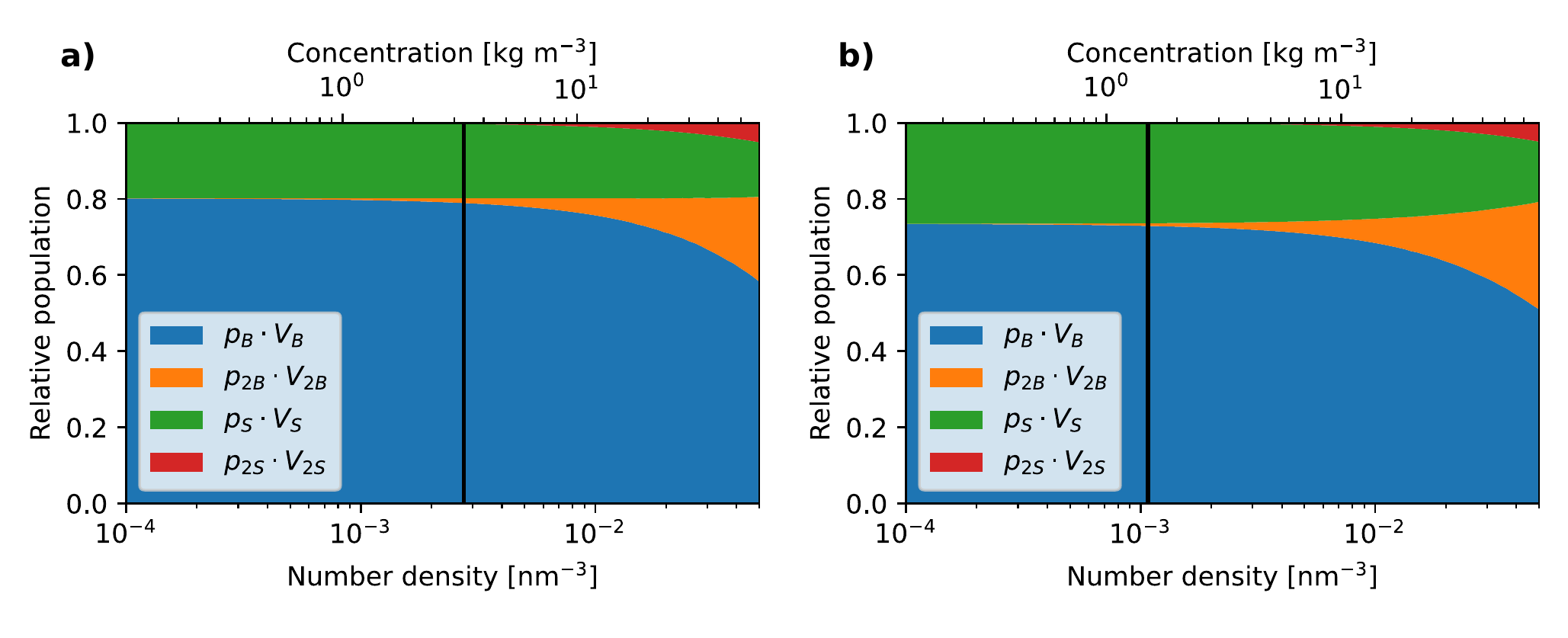}
    \caption{Concentration dependence of the population of the respective states, given as $p_x \cdot V_x$ for each state $x$.
    The concentration of the dimer states increases significantly only beyond the solubility limit indicated by the vertical full line (see text for details).}
    \label{fig:concentration-dependence}
\end{figure}

When investigating the effect of the dimers in a real system, the full weights $p_1 \cdot V_1$ and $p_2 \cdot V_2$ have to be compared.
Hereby, we can see, that the population of dimers is less than \SI{2}{\percent} for both C60 and C70 and all concentrations up to the solubility limit (cf. SI-Fig.~\ref{fig:concentration-dependence}).
The fact that the population increases only substantially beyond the limit of solubility (indicated by the solid line in SI-Fi.~\ref{fig:concentration-dependence}) is a clear indication, that close to the solubility limit, not only two particle interactions need to be accounted for, but also interactions between multiple particles.
According to theoretical predictions these are even more attractive than two particle interactions \cite{Prylutskyy2001} and thus expected to give rise to a way stronger increase of multimer population close to the solubility limit.

\section{Size of a fullerene}
Following Dresselhaus et al. \cite{Dresselhaus1996}, the size of a fullerene can be estimated by taking the radius of the nuclei structure and adding the approximate size of the electron orbital.
The latter was assumed to be the same as for graphene, i.e. half the graphitic interplanar distance of \SI{0.335}{\nm} \cite{Dresselhaus1996}.
The resulting radius for C60, when adding this to the carbon structure is $R_\textup{C60} = \SI{0.522}{\nm}$.
This is in excellent agreement with the radii obtained by various measurements for the intermolecular distances of crystalline C60 $R_\textup{C60, crystalline} \approx \SI{0.519}{\nm}$ \cite{Kraetschmer1990, Wilson1990, Goel2004}.

As C70 is anisotropic, determining the radius is slightly more complicated.
However, following the same procedure as for C60, the radius of the carbon structure is used and the size of the electron orbital is added.
We hereby use the average distance between the \gls{com} and the nuclei as the radius of the carbon structure.
This yields an average radius of $R_\textup{C70} = \SI{0.542}{\nm}$.

\section{Toluene diffusion in the pore}
\begin{figure}
    \centering
    \includegraphics[width=0.65\textwidth]{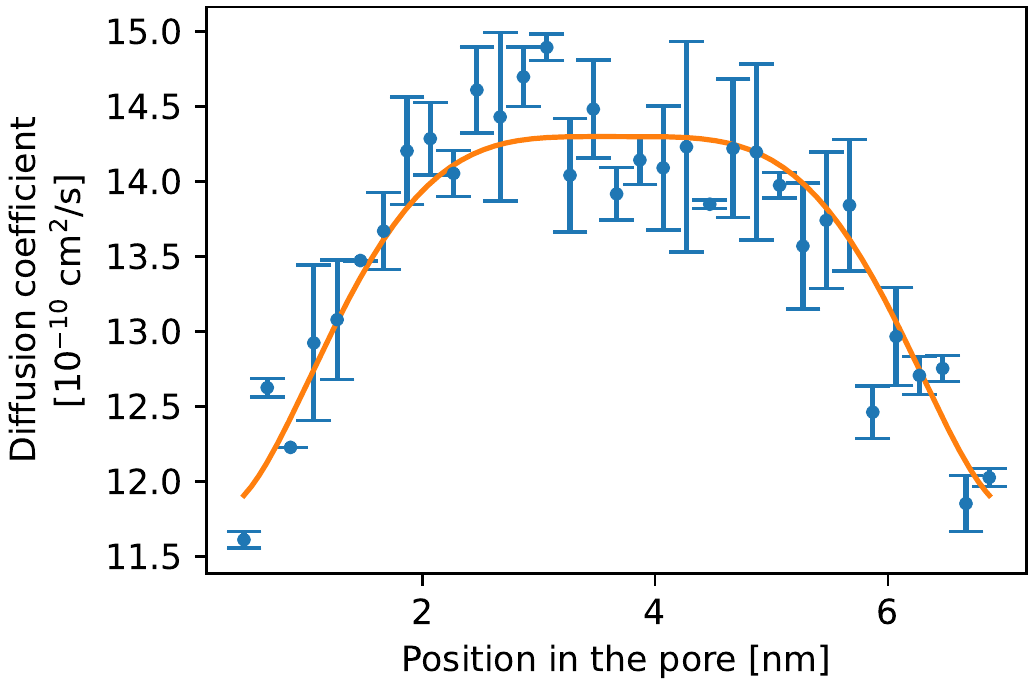}
    \caption{Lateral toluene diffusion within the pore. The standard deviation between the two equivalent directions is taken as an estimate for the statistical uncertainty of the obtained value. The solid line is a guide to the eye.}
    \label{fig:diffusion-toluene}
\end{figure}

The diffusion of toluene parallel to the pore wall is calculated as a function of the position within the pore (cf. SI-Fig.~\ref{fig:diffusion-toluene}).
The system is sliced in bins of \SI{0.2}{\nm} width and the mean squared displacement (MSD) is calculated for all molecules in the respective slab for the time that they stay in the respective slab.
The MSD is restricted to the lateral direction and calculated independently for both directions.
The standard deviation between the two equivalent directions is taken as an estimate for the statistical uncertainty of the obtained value.
The diffusion profile clearly resembles the increase in local toluene viscosity due to surface friction.

\end{document}